\documentclass[twocolumn,
superscriptaddress,amsmath, amssymb, amsfonts,preprintnumbers,aps,prd,
longbibliography,nofootinbib]{revtex4-2}

\usepackage{scalerel}
\usepackage{color}
\usepackage{amsmath,amsfonts,amssymb}
\usepackage{slashed}
\usepackage{latexsym,epsfig}
\usepackage{dsfont}
\usepackage{arydshln}
\usepackage{extarrows}
\usepackage{hyperref}
\usepackage{array}

\def\moth{\mathsurround=0pt}
\newdimen\zo \zo=0pt

\def\tick{\leaders\hrule height 0.5ex depth 0pt \hskip 0.5pt}
\def\upboxfill{$\moth \setbox\zo\hbox{\tick}%
  \hskip 3pt\hbox to 0pt{$\tick$\hss}\hrulefill \hbox to 7.5pt{$\tick$\hss}$}

\def\dtick{\leaders\hrule height .34pt depth 0.5ex \hskip 0.5pt}
\def\downboxfill{$\moth \setbox\zo\hbox{\dtick}%
  \hskip 2pt\hbox to 0pt{$\dtick$\hss}\hrulefill \hbox to 2pt{$\dtick$\hss}$}

\def\ov{\bar}

\def\bec{\begin{center}}
\def\ec{\end{center}}

\def\t{\tau}

\def\un{\underline}
\def\ov{\overline}

\def\cG{{\cal G}}

\def\cE{{\cal E}}

\def\cR{{\cal R}}

\def\cH{{\cal H}}

\def\cT{{\cal T}}

\def\nn{\nonumber}

 \def\det{{\rm det\,}}
\def\be{\begin{equation}}
\def\ee{\end{equation}}
\def\bea{\begin{eqnarray}}
\def\eea{\end{eqnarray}}
\def\ba{\begin{array}}
\def\ea{\end{array}}

\newcommand{\Exp}[1]{\operatorname{e}^{#1}}

\newcommand{\rmd}{{\mathrm{d}}}

\newcommand{\Lie}{\pounds}
\newcommand{\gLie}{\hat{\pounds}}

\newcommand{\GL}{\mathrm{GL}}
\newcommand{\OO}{\mathrm{O}}

\newcommand{\ubar}[1]{\,\underline{\phantom{h}}\hskip-6pt {#1}}

\newcommand{\covD}{{\raisebox{0.3ex}{\rotatebox[origin=c]{180}{$\triangle$}}}{}}

\begin{document}

\title{O$(D,D)$-covariant formulation of perfect and imperfect fluids in the double geometry}

\author{Eric Lescano} 
\email{elescano@irb.hr}
\affiliation{University of Wroclaw, Faculty of Physics and Astronomy, Maksa Borna 9, 50-204 Wroclaw,
Poland}

\affiliation{Division of Theoretical Physics, Rudjer Boskovic Institute, Bijenicka 54, 10000 Zagreb, Croatia}

\author{Nahuel Mirón-Granese}
\email{nahuelmg@fcaglp.unlp.edu.ar}
\affiliation{Consejo Nacional de Investigaciones Científicas y Técnicas (CONICET),
Godoy Cruz 2290, Ciudad de Buenos Aires C1425FQB, Argentina}

\affiliation{ Facultad de Ciencias Astronómicas y Geofísicas, Universidad Nacional de La Plata, Paseo del Bosque, La Plata B1900FWA, Buenos Aires, Argentina }

\affiliation{Universidad de Buenos Aires, Facultad de Ciencias Exactas y Naturales, Departamento de Física,\\Intendente Güiraldes 2160, Ciudad Universitaria, Ciudad de Buenos Aires C1428EGA, Argentina}

\author{Yuho Sakatani} 
\email{yuho@koto.kpu-m.ac.jp}
\affiliation{Department of Physics, Kyoto Prefectural University of Medicine,
1-5 Shimogamohangi-cho, Sakyo-ku, Kyoto 606-0823, Japan}


\begin{abstract}
We study generic matter coupled to a $D$-dimensional supergravity using a formulation of Double Field Theory (DFT), where all the fields are encoded in O$(D,D)$ multiplets. We study both the case when the matter comes from a variational principle, as well as the case where the matter comes from a statistical or thermodynamic approach. For the latter, we construct the distribution function for the perfect fluid and its entropy current, which is a conserved quantity. We then include general viscous and elastic terms in the generalized energy-momentum tensor which, in the general case, lead to entropy production. We consistently deform the conservation law of the generalized entropy current and identify a particular non-dissipative deformation. Using the generalized fluid model, we revisit the issue of non-covariance of perfect fluids under T-dualities and we show how to resolve it in our DFT model with matter.
\end{abstract}

\maketitle

\section{Introduction}
Symmetries and dualities play crucial roles in the understanding of physics. While symmetries are associated with invariances within a given framework, dualities hold a distinct place in identifying diverse formulations of the same physical theory becoming particularly useful when the theories are connected to string theory or its supergravity limit. Notably, dualities frequently unearth unexpected connections between seemingly distinct physical theories, enriching the landscape of string theory as well as allowing us to choose the most suitable framework to work with. In this work we study the effect of T-duality \cite{T-duality1}--\cite{T-duality3} on a convenient rewriting of the universal NS-NS supergravity backgrounds with generic matter content. We explore both scenarios: one in which the matter content comes from a Lagrangian defining a variational problem and another where the matter is defined through a statistical or effective approach. The former is suitable for a generic field theory on an arbitrary supergravity background, while the latter is adequate for describing effective fluid dynamics, such as in string cosmological scenarios, where the classical string sources are coupled to the standard supergravity background. In these cases one problem is that the fluid dynamics requires both perfect and imperfect contributions in order to realize the symmetry under T-dualities or $\OO(D,D)$ rotations\footnote{In section \ref{duality}, we study duality rotations for cosmological backgrounds, where the supergravity fields depend only on time. There, $\OO(D,D)$ is restricted to $\OO(D-1,D-1)$ associated with (Abelian) isometries along the spatial directions.} or, in other words, the energy-momentum tensor of the perfect fluid is mapped to the energy-momentum tensor of an imperfect fluid after the duality rotation. This issue was initially identified in \cite{GasVene} by using a cosmological ansatz. Although the problem was addressed at the supergravity level, it has direct consequences for the formulation of Double Field Theory (DFT) \cite{Siegel1}--\cite{DFT2}, since there is an apparent inconsistency in writing a perfect fluid in a fully $\OO(D,D)$-covariant way.

The standard DFT formalism\footnote{For reviews see \cite{ReviewDFT1}--\cite{ReviewDFT3} and the second lecture in \cite{ReviewDFT4}.} can be understood as a formal rewriting of the degrees of freedom of the theory in terms of multiplets of $\OO(D,D)$, which are defined on a double space whose local coordinates lie in the vector representation of the duality group. Alternatively, DFT can be viewed as the fundamental starting point for elaboration, with the low-energy limit of string theory emerging as a result of a strong constraint or section condition. Adopting this perspective, establishing the foundations of fluid mechanics and thermodynamics in DFT becomes a promising task which would allow us to understand the connection between the statistical description of strings and particles with the $\OO(D,D)$ symmetry.

The field content of DFT is given by a generalized metric and a generalized dilaton, and the dynamics can be determined by the generalized version of the Einstein equation and imposing the equation of motion for the generalized dilaton. Moreover, the effective matter dynamics can be defined through the conservation law of the generalized energy-momentum tensor which in turn is obtained from both a variational principle \cite{CurrentsDFT1,CurrentsDFT2} or using a statistical/kinetic approach in the double space \cite{EN1}. This last formulation is based on a double phase space configuration where the generalized Boltzmann equation is also considered. An attempt to construct hydrodynamics equations using a generalized velocity in the double space was realized in \cite{EN1}--\cite{EN3}, where an $\OO(D,D)$-covariant model of fluids was constructed considering a perfect fluid-scalar field correspondence. While this correspondence can be used at the supergravity level in order to construct the energy momentum-tensor for the perfect fluid from the scalar field dynamics, in the double space a second problem arises: by requiring the $\OO(D,D)$-covariance the authors in \cite{EN1}--\cite{EN3} could not reproduce the full energy-momentum tensor of perfect fluids in a generic supergravity background, because the pressure was related to the dilaton source as $p=\sigma/2$ in order not to spoil the generalized diffeomorphisms invariance. As we discuss later, this problem is closely related to the first problem (i.e., the non-covariance of the perfect fluid \cite{GasVene}) and therefore it is desirable to find an $\OO(D,D)$-covariant formulation that extends beyond the generalized scalar field-perfect fluid correspondence. 

In this work we address both issue simultaneously by proposing an  $\OO(D,D)$-covariant formulation for imperfect fluids. We begin with the construction of the generalized distribution function, akin to the Maxwell-Juttner distribution but in the double phase space. Using this function we prove that a double perfect fluid does not produce entropy. Then we propose a deformation of the generalized energy-momentum tensor of \cite{EN1}--\cite{EN3} using generalized strain tensors. Adhering to the second law of thermodynamics in the double space, we formulate a consistent model that generally leads to entropy production. 

The main conclusion is that the energy-momentum tensor ${T}^{\textrm{perfect}}_{mn}$ of perfect fluids discussed in \cite{GasVene} can be uplift to ${\cal T}^{\textrm{imperfect}}_{M N}$ of imperfect fluids in our $\OO(D,D)$-covariant model, where the generalized strain tensors are included. Since our model is $\OO(D,D)$ covariant by construction, the configuration of imperfect fluids is mapped to another configuration of imperfect fluids. As noted in \cite{GasVene} in the supergravity description, the dual configuration of ${T}^{\textrm{perfect}}_{mn}$ was identified as a configuration of viscous fluids.  The story can be summarized as
\bea
&{\cal T}^{\textrm{imperfect}}_{M N}  \xrightarrow{\OO(D,D)} {\cal T}^{\textrm{imperfect}}_{M N}& \nn \\ &\updownarrow \qquad \qquad \qquad \qquad \updownarrow&   \nn \\ 
&{T}^{\textrm{perfect}}_{mn}  \xrightarrow{\OO(D,D)}  {T}^{\textrm{viscous}}_{mn}& \,, \nn
\eea
where $M,N$ are $\OO(D,D)$ indices which run as $0,\dotsc,2D-1$, while $m,n=0,\dotsc,D-1$ are $\GL(D)$ indices.

Our main results are:
\begin{itemize}

\item We construct the distribution function for the perfect fluid in the double space. This function is derived through a suitable generalization of the Maxwell-Juttner distribution function \cite{Kinetic1}--\cite{Kinetic3} preserving all the constraints of the double phase space.

\item Using the generalized distribution function we give the formal construction of the entropy current and its conservation law. We prove that there is no entropy production for the double perfect fluid in the DFT formalism. 

\item We include new viscous (non-perfect) terms in the proposal \cite{EN1}--\cite{EN3} and we also propose a systematic way to write down the entropy production in these cases based on \cite{Y1,Y2}. 

\item We revisit the issue of the non-covariance of perfect fluid dynamics. It turns out that the $\OO (D,D)$-boost transformation of a perfect fluid configuration is identified as that of viscous fluids as it is shown in \cite{GasVene}. We highlight that, in the general case where the pressure and the energy density are generic, the energy-momentum tensor of the original configuration cannot be expressed in an $\OO(D,D)$-covariant way only in terms of the generalized metric, the generalized dilaton and the generalized velocity. We argue that the initial DFT configuration must inherently contain extra contributions. In this work, we define these additional terms as generalized strain tensors, introducing them as $\OO(D,D)$-covariant tensors. 
\end{itemize}

We begin the following section by discussing what kind of fluid scenarios have been described in the DFT and supergravity literature so far \cite{GasVene} \cite{EN1}--\cite{EN3} \cite{Meissner}--\cite{Brandenmatter}. 
See also \cite{cosmoDFTextra1}-\cite{cosmoDFTextra5} for other related works. 

\section{Coupling matter in Double Field Theory and supergravity}

\subsection{The Einstein equation in DFT and supergravity}

DFT is a T-duality-invariant reformulation of the low-energy limit of string theory. This theory is defined on a double space and all the fields and parameters transform covariantly under $\OO(D,D)$. The invariant group metric is given by
\begin{align}
{\eta}_{{M N}} = \left(\begin{matrix}0&\delta_m^n\\ 
\delta^m_n&0 \end{matrix}\right)\,, \label{etaintro}
\end{align}
where $m,n=0,\dotsc,D-1$. The metric (\ref{etaintro}) is used to raise and lower the $\OO(D,D)$ indices $M,N, \cdots$. Infinitesimal generalized diffeomorphisms are consistently defined with respect to this metric and then generalized vectors and tensors can be constructed \cite{RiemannDFT1}--\cite{RiemannDFT3}. Within this framework one can write the generalized Einstein equation as \cite{ParkCosmo1}--\cite{ParkCosmo2},
\begin{align}
\mathcal{G}_{MN}=\mathcal{T}_{MN}\label{einstein} \,,
\end{align}
where the generalized Einstein tensor is defined by
\begin{align}
\mathcal{G}_{MN}=\mathcal{R}_{MN}-\frac12\,\mathcal{R}\,{\cal H}_{MN} \,,
\end{align}
and $\mathcal{R}_{MN}=\mathcal{R}_{MN}({\cal H},d)$, $\mathcal{R}={\cal R}({\cal H},d)$ are the generalized Ricci tensor and scalar, respectively. The generalized energy-momentum tensor is given by
\begin{align}
\mathcal{T}_{MN}=\hat{\mathcal{T}}_{MN}-\frac12 \,\cT\,{\cal H}_{MN}\,,\label{tmntotal}
\end{align}
where both ${\cal R}_{MN}$ and $\hat{\cT}_{MN}$ contain only mixed components with respect to the DFT projectors $P_{M N} = \frac12(\eta - {\cal H})_{M N}$ and $ \ov P_{M N} = \frac12(\eta + {\cal H})_{M N}$\footnote{An arbitrary vector can be projected as $W^{M} = P^{M}{}_{N} W^{N} + \ov P^{M}{}_{N} W^{N} = W^{\un M} + W^{\ov M}$. By ``mixed components'' we mean that we have, for example, ${\cal R}_{MN}={\cal R}_{\un M\ov N}+{\cal R}_{\ov M\un N}$.}. In consequence, (\ref{einstein}) reads
\begin{align}
\mathcal{R}_{MN}&= \hat{\mathcal{T}}_{MN}\,,\label{rmneq}
\\
\mathcal{R}&= \mathcal{T}\label{req}\,.
\end{align}

A crucial consistency equation of the DFT formulation is given by the strong constraint 
\begin{align}
\partial_{M} (\partial^{M} \star) = 0 \, \quad (\partial_{M} \star) (\partial^{M} \star) = 0 \,,
\label{SC}
\end{align}
where $\partial_{M}=(\partial_{m},\tilde{\partial}^m)$ refers to derivatives with respect to the ordinary/dual coordinates, respectively, and $\star$ represents any combination of generalized fields and/or gauge parameters. The usual solution to (\ref{SC}) in order to recover the supergravity framework is $\tilde{\partial}^m=0$\,. The generalized metric/dilaton encodes the field content of the universal NS-NS sector,
\begin{align}
 \cH_{MN} & = \begin{pmatrix}
 \delta_m^p & B_{mp} \\
 0 & \delta_p^m
 \end{pmatrix}
 \begin{pmatrix}
 g_{pq} & 0 \\
 0 & g^{pq}
 \end{pmatrix} 
 \begin{pmatrix}
 \delta_n^q & 0 \\
 -B_{qn} & \delta_q^n
 \end{pmatrix},\\
 \Exp{-2d} & = \Exp{-2\phi}\sqrt{-g}\,.
\label{param}
\end{align}

Then the generalized Ricci tensor and the generalized Ricci scalar become
\begin{align}
 \cR_{MN} &=\begin{pmatrix}
 \delta_m^p & B_{mp} \\
 0 & \delta_p^m
 \end{pmatrix}
 \begin{pmatrix}
 - s_{(pq)} & - g_{pr}\,s^{[rq]} \\
 s^{[pr]}\,g_{rq} & s^{(pq)}
 \end{pmatrix} 
 \begin{pmatrix}
 \delta_n^q & 0 \\
 -B_{qn} & \delta_q^n
 \end{pmatrix},
\\
 \cR &= R + 4\,\covD^m \partial_m \phi - 4\,\left(\partial \phi\right)^2 - \tfrac{1}{12}\,H^2 \,,
\end{align}
where $\covD_m$ denotes the ordinary (torsionless) covariant derivative, $H_{mpq}=3\,\partial_{[m} B_{pq]}$, $H^2=H_{mpq} H^{mpq}$, and
\begin{align}
 s_{(mn)} & = R_{mn}-\tfrac{1}{4}\,H_{mpq}\,H_n{}^{pq} + 2\,\covD_m \covD_n \phi \,,
\\
 s_{[mn]} & = - \tfrac{1}{2}\Exp{2\phi} \covD^p \bigl(\Exp{-2\phi} H_{pmn}\bigr) \,, 
\end{align}
where $R$ and $R_{mn}$ are the usual Ricci scalar and Ricci tensor. In addition, the generalized energy-momentum tensor can be generically parametrized
as
\begin{align}
\hat{\mathcal{T}}_{MN}=\begin{pmatrix}\delta^p_m&B_{mp}\\0&\delta^m_p\end{pmatrix}\begin{pmatrix}-t_{(pq)}&-g_{pr}t^{[rq]}\\t^{[pr]}g_{rq}&t^{(pq)}\end{pmatrix}\begin{pmatrix}\delta^q_n&0\\-B_{qn}&\delta^n_q\end{pmatrix} .\nn\\\label{tmnmixed}
\end{align}
Then (\ref{rmneq}) and (\ref{req}) give,
\begin{align}
R_{mn}-\frac14 H_{mpq} H_n{}^{pq}+2\,\covD_m\partial_n\phi&= t_{(mn)}\,,\label{mixed1}\\
-\frac12\,\covD^p H_{pmn} +\partial_p\phi\, H^p{}_{mn}&= t_{[mn]}\,,\label{mixed2}\\
R+4\,\covD^m \partial_m\phi-4\left(\partial \phi\right)^2-\tfrac{1}{12} H^2&= \cT\label{rt}\,.
\end{align}
Under this parameterization and $\tilde{\partial}^m=0$, the conservation law $\nabla_M\cT^{MN}=0$ reads
\begin{align}
 &\covD^n \left[\Exp{-2\phi}\,(t_{(nm)}-\tfrac{1}{2}\,\cT\,g_{nm})\right] = \Exp{-2\phi}\,\cT\,\partial_m\phi\,,
\\
 &\covD_n (\Exp{-2\phi}\,t^{[nm]}) =0\,.
\end{align}
The equations of motion for the metric, the $B$-field, and the dilaton in the presence of matter fields can be described as
\begin{align}
 & G_{mn} + 2\,\covD_m \covD_n \phi+ 2\,g_{m n}(\covD\phi)^2 -2\,g_{m n}\covD^2\phi \nn \\
 & + \tfrac{1}{24}\, g_{m n} H^2 - \tfrac{1}{4} H_{mpq} H_{n}{}^{pq} = \Exp{2\phi}\,T_{mn} \,, 
\label{emt}
\\
 &-\tfrac12\,\covD_p H^{pmn}+(\covD_p \phi)H^{pmn}=2\Exp{2\phi}J^{mn}\,,
\label{jeq}
\\
 & R - 4\,(\partial \phi)^2 - \tfrac{1}{12}\,H^2 + 4 \covD^{m} \covD_{m} \phi =- \Exp{2\phi}\,\sigma\,,
\label{dilatoneq}
\end{align}
where $G_{mn}$ is the Einstein tensor, $G_{mn}=R_{mn} - \frac12 R\, g_{mn}$ and the exponential factor $\Exp{2\phi}$ behaves as an effective gravitational coupling. 
Comparing Eqs.~(\ref{mixed1})--(\ref{rt}) with Eqs.~(\ref{emt})--(\ref{dilatoneq}) one finds
\begin{align}
t_{(mn)}&= \Exp{2\phi}\left[T_{mn}-\frac{\sigma}{2}\,g_{mn}\right] ,\label{t(mn)}\\
\cT &= -\Exp{2\phi} \sigma\,,\label{calt}\\
t_{[mn]}&= 2 \Exp{2\phi} J_{mn}\label{t[mn]}\,,
\end{align}
and the conservation law is
\begin{align}
 \covD^m T_{mn} = -\sigma\,\partial_n\phi\,,
\quad
 \covD^m J_{mn} =0\,.
\end{align}

\subsection{The variational principle in DFT and supergravity}

So far, the energy-momentum tensor $\mathcal{T}_{MN}$ can be arbitrary, but let us consider a case where the dynamics of the matter fields can be described by the action principle. 
We denote the action as $S=S_0+S_{\rm mat}$ with 
\begin{align}
 S_0 = \frac{1}{2}\,\int \rmd^Dx\, \rmd^D\tilde x \,\Exp{-2d}\,\cR({\cal H},d)\,,
\end{align}
and the matter action given by
\begin{align}
 S_{\rm mat} = \int \rmd^Dx\, \rmd^D\tilde x\,\Exp{-2d}\,L_{\rm mat} \,.
\end{align}
The generalized equations of motion and the generalized energy-momentum tensor can be read from the variation
\begin{align}
\!\!\!
 \delta S_{\rm mat} = \Exp{-2d} \left(\tfrac{1}{4}\,\hat{\mathcal{T}}_{MN}\,\delta \cH^{MN} + \cT\,\delta d \right)
 + \frac{\delta S_{\rm mat}}{\delta \Psi}\, \delta \Psi\,,
\end{align}
where $\Psi$ collectively denotes the matter fields.

Using the same parametrization as in the previous section one obtains the standard NS-NS supergravity formulation in which
\begin{align}
\!\!\! S_{0} = \frac12\int \rmd^Dx\,\Exp{-2\phi}\sqrt{-g}\left[R+4\,(\partial \phi)^2 - \frac{1}{12}\,H^2 \right],
\end{align}
and
\begin{align}
 S_{\rm mat} = \int \rmd^Dx \,\Exp{-2\phi}\sqrt{-g}\,L_{\rm mat}\,,\label{matteraction}
\end{align}
while the matter sources are given by
\begin{align}
 T_{mn}&= -\frac{2}{\sqrt{-g}} \frac{\delta S_{\rm mat}}{\delta g^{mn}}\,,\label{deftmunu}
\\
 \sigma &= -\frac{1}{\sqrt{-g}}\,\frac{\delta S_{\rm mat}}{\delta \phi}\,,\label{defsigma}
\\
J^{mn} &= - \frac{2}{\sqrt{-g}}\frac{\delta S_{\rm mat}}{\delta B_{mn}}\,.\label{defj}
\end{align}

In the case where the matter fields $\Psi$ minimally couple to the metric and the dilaton \cite{quintin}, the matter Lagrangian $L_{\rm mat}$ in (\ref{matteraction}) only depends on $g_{mn}$, $B_{mn}$ and $\Psi$, and not on $\phi$. Then the matter sources for this action become
\begin{align}
T_{mn}&= \Exp{-2\phi}\left[g_{mn}\,L_{\rm mat} - 2\,\frac{\delta L_{\rm mat}}{\delta g^{mn}} \right] \,, \\
\sigma &= 2 \Exp{-2\phi}\,L_{\rm mat} \,, 
\label{sigmaminimally}\\
J^{mn}&= -2 \Exp{-2\phi}\,\frac{\delta L_{\rm mat}}{\delta B_{mn}}\,.
\end{align}
From Eq.~(\ref{sigmaminimally}) one observes that, for this kind of framework, a vanishing dilaton source ($\sigma=0$) is not possible.

Among this kind of minimally coupled matters in supergravity, we analyze two specific cases. On the one hand we consider that the matter is given by a scalar field for which its sources read
\begin{align}
T_{mn}&= \Exp{-2\phi}\left[g_{mn} \left(-\frac12\,\covD_p\Phi\,\covD^p\Phi-V(\Phi)\right)+\partial_m \Phi\,\partial_n\Phi\right],
\\
 \sigma &= 2\,\Exp{-2\phi}\left(-\frac12\,\covD_m\Phi\,\covD^m\Phi-V(\Phi)\right),
\\
J_{mn}&= 0\,.
\end{align}
Since it is possible to establish a formal correspondence with the perfect fluid $T_{mn}=(e+p)\,u_mu_n+p\,g_{mn}$ \cite{EN3}, we can read the energy density and pressure under the identifications $u_m \propto \covD_m\Phi$ and
\begin{align}
 e&= \Exp{-2\phi}\,\Bigl[-\frac12\,\covD_m\Phi\,\covD^m\Phi+V(\Phi)\Bigr]\,,
\\
 p&= \Exp{-2\phi}\,\Bigl[-\frac12\,\covD_m\Phi\,\covD^m\Phi-V(\Phi)\Bigr]\,.
\end{align}
Interestingly enough we observe that the supergravity matter Lagrangian can be given by $L_{\textrm{mat}}=\Exp{2\phi} p$ and, furthermore, the dilaton source $\sigma$ fixes its value according to $\sigma=2\,p$. These relations explain how the perfect fluid dynamics can be constructed using a formal correspondence with the scalar field dynamics. 

On the other hand we consider minimally coupled matter within a cosmological ansatz in which all the variables depend only on time making the theory $\OO(D-1,D-1)$-invariant and we assume that the energy-momentum tensor corresponds to an isotropic perfect fluid $T^m{}_n={\rm diag}(-e,\,p\,\delta^i{}_j)$. Then we can use the following definition of the pressure, regardless the specific matter content,
\begin{align}
 p&= \frac{-2}{\sqrt{-g}}\frac{g^{ij}}{D-1}\frac{\delta S_{\rm mat}}{\delta g^{ij}}
\nn\\
 &= \frac{-2}{\sqrt{-g}}\frac{g^{ij}}{D-1}\left[\frac{\delta S_{\rm mat}}{\delta g^{ij}}\bigg|_{d}+\frac{\delta S_{\rm mat}}{\delta d}\bigg|_{\cal H}\frac{\delta d}{\delta g^{ij}}\right]
\nn\\
 &= p_{\rm cov}+\frac\sigma 2\,,
\end{align}
where the spatial indices are $i,j=1,\dotsc,D-1$ and $p_{\rm cov}$ is what is called the $\OO(D-1,D-1)$-covariant pressure in \cite{quintin}. Naturally for this cosmological ansatz the matter Lagrangian does not depend on the spatial metric, thus $p_{\rm cov}=0$ and again $\sigma=2\,p$.

In order to recover the previous cases from the DFT framework, let us consider the most general expression for the generalized energy-momentum tensor $\mathcal{T}_{MN}$ depending on the generalized velocity vector fields $U^M$ and the generalized metric $\cH_{MN}$ \cite{EN2},
\begin{align}
 \mathcal{T}_{MN}=a\left( U_{\ov { M}}\, U_{\un { N}}+ U_{\ov { N}}\, U_{\un { M}}\right)+b\,{\cal H}_{ MN}\,,
\label{proposaltmn}
\end{align}
where the coefficients $a$ and $b$ can be, in principle, arbitrary generalized scalars. Here we stress that the proposal (\ref{proposaltmn}) is valid for a generic background and not just for a cosmological ansatz.

The decomposition of the Eq.~(\ref{tmntotal}) gives
\begin{align}
\hat{\cT}_{MN} & = a\left( U_{\ov { M}}\, U_{\un { N}}+ U_{\ov { N}}\, U_{\un { M}}\right) \,, 
\label{hatcT-p}\\
\mathcal{T} & = -2\,b \,,
\end{align}
and after imposing the cosmological ansatz one is forced to described cosmological scenarios with
\begin{align}
\label{etoymodel}
e&= \Exp{-2\phi}\,\left(\frac{a}2-b\right)\,,\\ \label{ptoymodel}
 p&= \sigma/2=\Exp{-2\phi}\,b\,,\\
 J^{mn}&= 0\,.
\label{Jtoymodel}
\end{align}
In terms of Eqs.~(\ref{t(mn)})--(\ref{t[mn]}) and by virtue of a generalized correspondence with a generalized scalar field \cite{EN3} one finally obtain
\begin{align}
 t_{(mn)}&= (\mathsf{e}+ \mathsf{p})\,u_m\,u_n\,,\\
 \cT &= -2\,\mathsf{p}\,,\\
 t_{[mn]}&= 0\,,
\end{align}
where we have defined $\mathsf{e}=\Exp{2\phi} e$ and $\mathsf{p}=\Exp{2\phi} p$\,, which behave as generalized scalars fields in DFT. 
In this case, the energy-momentum reads
\begin{align}
 \cT_{MN} = 2\,(\mathsf{e}+\mathsf{p})\,\bigl(U_{\ubar{M}}\,U_{\bar{N}} + U_{\bar{M}}\,U_{\ubar{N}}\bigr) + \mathsf{p}\,\cH_{MN} \,.
\label{Gemt}
\end{align}
and it enables us to recover a family of string cosmologies as discussed in \cite{EN2}. In order to go beyond the constraints imposed by (\ref{etoymodel})--(\ref{Jtoymodel}), one possibility is to introduce additional variables in ${\cal T}_{M N}$ which means going beyond the double perfect fluid defined through the correspondence between this fluid and the generalized scalar field dynamics \cite{EN3}.

In any case a statistical interpretation in terms of a double kinetic theory may be considered \cite{EN1}. Since the generalized distribution function for the perfect fluid was not constructed, in Section \ref{statistics} we use the phase space formulation of DFT to construct a suitable proposal for it with a vanishing entropy production. Afterwards, in Section \ref{imperfectfluids} we study non-perfect contributions and its non-negative entropy production considering additional variables and news terms directly in the energy-momentum tensor.

\section{Fluid Statistics in the double geometry}\label{statistics}

For perfect fluids, the right hand side of (\ref{einstein}) can be constructed in the phase space framework of DFT \cite{EN1}. Momentum coordinates of the form ${\cal P}^{M}=(\tilde p^{m},\,p_{m})$ have to be included and the generalized Lie derivative is consistently deformed in order to define diffeomorphism invariance.

The generalized Boltzmann equation is given by
\begin{align}
{\cal P}^{M} {\cal D}_{M} F = {\cal C}[F] \,,
\label{BoltzmannDFT}
\end{align}
where ${\cal C}[F]$ is the generalized collision term and the operator ${\cal D}_M$ is
\begin{align}
{\cal D}_{M} = D_{M} - {\cal U}_{M}\,,
\end{align}
with ${\cal U}_{M}= 2\, \partial_M d$ and ${D}_{M} = \nabla_{M} - \Gamma_{M N}{}^{Q} {\cal P}^{N} \frac{\partial}{\partial {\cal P}^{Q}}$. This last derivative is the covariant derivative in the phase space, which can be easily constructed demanding that the derivative of the phase space scalars transform correctly. The dilatonic contribution in (\ref{BoltzmannDFT}) is due to the fact that the generalized distribution function is a phase-space density scalar, so that the integration of this quantity in the double space produces double space-time tensors (See for example the RHS of equations (\ref{EMDFT}) and (\ref{SDFT}) which transform as generalized tensors). At this point we can consider an equilibrium state such that ${\cal C}[F_{\rm eq}]=0$ and we propose the following ansatz for the generalized equilibrium distribution function,
\begin{align}
F_{\rm eq}(X,{\cal P})= \Exp{2d - {\cal P}^{M} {\cal H}_{M N} \beta^{N}}
\label{MBDFT}
\end{align}
where $\beta_{N} = U_{N}/T$ is a generalized Killing vector, i.e.,
\begin{align}
2 \nabla_{(M|} \beta^{P} {\cal H}_{P| N)} - 2 \nabla^{P} \beta_{(M|} {\cal H}_{P| N)} = 0 \,, 
\end{align}
and $T$ is the $\OO(D,D)$ temperature. Replacing the previous generalization of the Maxwell-Juttner distribution function in the generalized version of the Boltzmann equation we obtain
\begin{align}
{\cal P}^{M} \nabla_{M} \beta_{P} {\cal H}^{P}{}_{Q} {\cal P}^{Q} = 0 \,, 
\label{Boltzmann}
\end{align}
which implies
\begin{align}
{\cal P}^{M} \nabla^{P} \beta_{M} {\cal H}_{P N} {\cal P}^{N} = 0 \,,
\label{DFTK}
\end{align}
through the generalized Killing equation of $\beta_{M}$. On the other hand, the transfer equation of the double kinetic theory is given by \cite{EN1}
\begin{align}
 & \nabla_{N}\Big[\int \Psi^{M} {\cal P}^{N} F_{\rm eq} \Exp{-2d} \rmd^{2D}{\cal P} \Big] \nn \\
 & - \int F_{\rm eq} {\cal P}^{N} {D}_{N} \Psi^{M} \Exp{-2d} \rmd^{2D}{\cal P} = 0 \,,
\label{lawDFT}
\end{align}
with $\Psi^{M}$ an arbitrary phase-space covariant object. Using
\begin{align}
\Psi_{M} = {\cal P}_{M} \,, 
\end{align}
we can formally define the generalized energy-momentum tensor as
\begin{align}
\label{EMDFT}
\cT^{M N} = \int \Exp{-2d} {\cal P}^{M} {\cal P}^{N} F_{\rm eq} \,\rmd^D{\cal P} \,,
\end{align}
which, after imposing the generalized version of the scalar field-perfect fluid correspondence \cite{EN3}, should be related to (\ref{Gemt}). On the other hand by choosing
\begin{align}
\Psi = \ln{(\Exp{2d} F_{\rm eq}}) \,,
\end{align}
in (\ref{lawDFT}) we can formally define the generalized entropy current as
\begin{align}
S^{M} = \int \Exp{-2d} {\cal P}^{M} F_{\rm eq} \ln{(\Exp{-2d} F_{\rm eq})} \,\rmd^D{\cal P} \,.
\label{SDFT}
\end{align}
In a general system the conservation equation for the generalized entropy current is given by
\begin{align}
\nabla_{N} S^{N} = \int F_{\rm eq} {\cal P}^{N} {D}_{N} \ln{(\Exp{-2d} F_{\rm eq})} \Exp{-2d} \rmd^{2D}{\cal P} \,,
\end{align}
which we can use to define the second law of thermodynamics for statistical matter coupled to a generic DFT background. Furthermore, when we inspect the particular case of the perfect fluid in the double space we find the expected conservation law
\begin{align}
\nabla_{N} S^{N} & = - \int \Exp{- {\cal P}^{M} {\cal H}_{M N} \beta^{N}} {\cal P}^{N} {\cal P}^{R} {\cal H}_{R S} {D}_{N} {\beta^{S}} \rmd^{2D}{\cal P} \, \nn\\
 & = - \int \Exp{- {\cal P}^{M} {\cal H}_{M N} \beta^{N}} {\cal P}^{N} {\cal P}^{R} {\cal H}_{R S} {\nabla}_{N} {\beta^{S}} \rmd^{2D}{\cal P} \nn \\
 & = 0 \,.
\end{align}
In the last step we use the independence between $\beta_{M}$ and ${\cal P}_{M}$ in order to transform the phase space covariant derivative into the ordinary one, and also (\ref{Boltzmann}) to show that the generalized current of entropy is conserved in this case.

\section{Imperfect fluids in the double geometry}\label{imperfectfluids}

The main goal of this work is to construct imperfect contributions in the generalized energy-momentum tensor given in (\ref{Gemt}). The most canonical way to do so is to consider a deformation of the generalized distribution function (\ref{MBDFT}). In this way the quantities (\ref{EMDFT}) and (\ref{SDFT}) will be deformed in order to include imperfect terms/effects. Here we take a different way based on thermodynamics, and construct the energy-momentum tensor $\cT^{MN}$ which contains imperfect contributions coming from a viscoelastic model \cite{Y1,Y2}. 

\subsection{The general case with entropy production}

We introduce additional thermodynamic variables ${\cE}_{MN}$ and $\varepsilon$ which represent generalized strain tensors. They describe the difference between the shape of a material before and after elastic deformation. We assume that the strain is small and then the strain tensor ${\cE}_{MN}$ corresponds to an infinitesimal variation of the generalized metric $\delta\cH_{MN}$\,. Since an infinitesimal variation of the generalized metric $\cH_{MN}$ only has the mixed components,\footnote{This can be shown by using $\delta P^{M}{}_{Q}\,\ov{P}^{Q}{}_{N}+P^{M}{}_{Q}\,\delta\ov{P}^{Q}{}_{N}=0$, $\delta P_{MN}=-\frac{1}{2}\,\delta\cH_{MN}$, and $\delta \ov{P}_{MN}=\frac{1}{2}\,\delta\cH_{MN}$\,.} the strain tensor ${\cE}_{MN}$ satisfies the property
\begin{align}
 {\cE}_{MN} = {\cE}_{(MN)} = {\cE}_{\bar{M}\ubar{N}} +{\cE}_{\ubar{M}\bar{N}}\,.
\label{projE}
\end{align}
For simplicity, we suppose that the strain tensor is orthogonal to the fluid velocity ${\cE}_{MN}\,\cH^{MN}\,U_N=0$\,. 
In DFT, the volume factor is contained in $\Exp{-2d}$, and the strain tensor associated with the bulk compression is introduced as a generalized scalar field $\varepsilon$\,.

Now the energy-momentum tensor $\cT^{MN}$ contains corrections so we propose
\begin{align}
 \cT^{MN} & = 2\,(\mathsf{e}+\mathsf{p})\,\bigl(U^{\ubar{M}}\,U^{\bar{N}} + U^{\bar{M}}\,U^{\ubar{N}}\bigr) + \mathsf{p}\,\cH^{MN} \nn \\
 & + \hat{\tau}^{MN} + \tau\,\Delta^{MN}\,,
\end{align}
where $\Delta^{MN}= 2\,(U^{\ubar{M}}\,U^{\bar{N}} + U^{\bar{M}}\,U^{\ubar{N}}) + \cH^{MN}$\,, and the corrections added in the second line are supposed to be spatial tensors; $\hat{\tau}^{MP}\,\cH_{PQ}\,U^Q=0$\,. 
We also suppose that $\hat{\tau}^{MN}$ only have mixed projections. In the following, we determine the explicit form of $\hat{\tau}^{MN}$ and $\tau$ from a thermodynamic point of view in the double space. Let us assume that the local thermodynamic equilibrium is realized, and we introduce the entropy density as
\begin{align}
 \tilde{s}(\tilde{P}_M,\,\cH_{MN},\,\Exp{-2d},\,{\cE}_{MN},\,\varepsilon)\,.
\end{align}
Here, the $\tilde{s}=\Exp{-2d}s$ and $\tilde{P}_M=\Exp{-2d}P_M$ contain the volume factor $\Exp{-2d}$ and $P_M=\mathsf{e}\,V_M$ ($V_M=\cH_{MN}\,U^N$) is the energy-momentum vector. We also assume that the configuration is close to the equilibrium and then the entropy density is a T-duality invariant generalization of the one in \cite{Y1,Y2},
\begin{align}
 \tilde{s} &= \tilde{s}_0(\tilde{P}_M,\,\cH_{MN},\,\Exp{-2d})
\nn\\
 &\quad + \frac{\lambda\Exp{-2d}}{2\,T}\,\,\cE^{MN}\,\cE_{MN}
  - \frac{\gamma \,\Exp{-2d}}{2\,T}\, \varepsilon^2 \,,
\end{align}
where $\lambda\geq 0$ and $\gamma \geq 0$\,. Its infinitesimal variation (under our assumption that the strain is small) becomes
\begin{align}
 \delta \tilde{s} = \delta \tilde{s}_0
+ \frac{\lambda\Exp{-2d}}{T}\, \cE^{MN}\,\delta \cE_{MN} 
   - \frac{\gamma\Exp{-2d}}{T}\, \varepsilon \,\delta\varepsilon \,,
\label{eq:ds-viscoelastic}
\end{align}
where the first term can be expanded as
\begin{align}
 \delta \tilde{s}_0 = \frac{\delta \tilde{s}_0}{\delta \tilde{P}_M}\,\delta \tilde{P}_M + \frac{\delta \tilde{s}_0}{\delta \cH_{MN}}\,\delta\cH_{MN} + \frac{\delta \tilde{s}_0}{\delta \Exp{-2d}}\,\delta \Exp{-2d}\,.
\label{eq:ds-viscoelastic0}
\end{align}
At this point we would like to rewrite the previous equation as $T\,\delta S \sim \delta E + p\,\delta V$. 
As we show later in Eq.~(\ref{TdS}), this can be realized by making the identification
\begin{align}
 \frac{\delta \tilde{s}_0}{\delta \tilde{P}_M} &= - \frac{U^{M}}{T}\,,\qquad
 \frac{\delta \tilde{s}_0}{\delta \Exp{-2d}} = \frac{\mathsf{p}}{T} \,,
\\
 \frac{\delta \tilde{s}_0}{\delta \cH_{MN}} &= \frac{\Exp{-2d}}{2\,T}\, (\mathsf{e}+\mathsf{p})\,\bigl(U^{\ubar{M}}\,U^{\bar{N}} + U^{\bar{M}}\,U^{\ubar{N}}\bigr) \,.\nn
\end{align}
We then consider the variation $\delta$ in Eq.~(\ref{eq:ds-viscoelastic}) as the generalized Lie derivative along the flow $U^M$, i.e., $\delta=\gLie_U$\,. 
We relate the entropy density with the entropy current as $S^M=s\,U^M$
and then,
\begin{align}
 \delta \tilde{s} &= \gLie_U \tilde{s} = U^M\,\partial_M\bigl(\Exp{-2d} s\bigr) + \partial_M U^M\,\Exp{-2d} s
\nn\\
 &= \Exp{-2d} \nabla_M S^M\,.
\end{align}
If we now inspect the RHS of (\ref{eq:ds-viscoelastic0}), the term including $\delta \cH_{MN}$ vanish due to the identity $U^M\,U^N\, \gLie_U \cH_{MN}=0$, and we also find that the term including $\delta\tilde{P}_M$ becomes
\begin{align}
 - \frac{U^{M}}{T}\,\gLie_U \tilde{P}_M &= -\frac{\Exp{-2d}\,V^M}{T}\,\nabla_N \bigl( \mathsf{e}\,U_M\, U^N \bigr) 
\nn\\
 &= -\frac{\Exp{-2d}\,V_M}{T}\,\nabla_N \bigl( {\cal T}^{MN} - T_{\textrm{s}}^{MN} \bigr)\,,
\end{align}
where we have defined the spatial part of the generalized energy-momentum tensor as
\begin{align}
 T_{\textrm{s}}^{MN} &= \bigl(\mathsf{p}+\tau\bigr)\,\Delta^{MN} + \hat{\tau}^{MN}\,,
\end{align}
which satisfies $V_M\,T_{\textrm{s}}^{MN}=0$\,. 
Then using the conservation law for the full generalized energy-momentum tensor, $\nabla_N {\cal T}^{MN}=0$\,, one gets
\begin{align}
 &- \frac{U^{M}}{T}\,\gLie_U \tilde{P}_M = - \frac{\Exp{-2d}}{T}\,T_{\textrm{s}}^{MN}\,\nabla_N V_M
\nn\\
 &= - \frac{\Exp{-2d}\,}{4\,T}\,\hat{\tau}^{MN}\,\gLie_U\cH_{MN} 
  - \frac{\Exp{-2d}\,}{T}\,(\mathsf{p}+\tau)\,\nabla_M U^M \,.
\end{align}
Finally, it is useful to notice that
\begin{align}
 \frac{\mathsf{p}}{T} \,\gLie_U \Exp{-2d} = \Exp{-2d}\,\frac{\mathsf{p}}{T}\,\nabla_M U^M \,.
\end{align}
Combining the previous expressions, we find
\begin{align}
 T\,\nabla_M S^M &= - \frac{1}{4}\,\hat{\tau}^{MN} \,\gLie_U\cH_{MN} + 2\,\tau \,\gLie_Ud
\nn\\
 &\quad + \lambda \, \cE^{MN}\,\gLie_U\cE_{MN} - \gamma \,\gLie_U\varepsilon \,\varepsilon \,.
\label{EntropyProd}
\end{align}
The equation (\ref{EntropyProd}) is another way to prove that the perfect fluid does not produce entropy because all the quantities in the RHS of (\ref{EntropyProd}) depend on the imperfect contributions. This derivation also dictates how to measure the entropy production for a given $\hat \tau_{M N}$, $\tau$, $\cE_{M N}$ and $\varepsilon$. Now we assume that $\hat{\tau}^{MN}$ and $\tau$ are given by linear combinations of $\OO(D,D)$-covariant quantities, all of which should vanish at the equilibrium. 
Then, to make the right-hand side of Eq.~(\ref{EntropyProd}) non-negative, i.e. $\nabla_M S^M\geq0$, we require
\begin{align}
\begin{pmatrix} 
 -\lambda \, [\gLie_U\cE]_{MN} \\ \frac{1}{2}\,\hat{\tau}_{MN}
\end{pmatrix}
 &= (\boldsymbol{\cG}+\boldsymbol{\eta})
 \begin{pmatrix} 
 \cE_{MN} \\ \frac{1}{2}\, \gLie_U\cH_{MN}
\end{pmatrix},
\label{require1}
\\
 \begin{pmatrix}
 - \gamma \gLie_U\varepsilon \\
 \tau
\end{pmatrix}
 &= (\boldsymbol{\cal K}+\boldsymbol{\zeta})
 \begin{pmatrix}
 \varepsilon \\ 2\,\gLie_U d
\end{pmatrix},
\label{require2}
\end{align}
where the bracket in $[\gLie_U\cE]_{MN}$ denotes the projection into mixed components, and
\begin{alignat}{2}
 \boldsymbol{\cG} &= \begin{pmatrix} 0 & \cG \\ -\cG & 0 \end{pmatrix},\quad&
 \boldsymbol{\cal K} &= \begin{pmatrix} 0 & {\cal K} \\ -{\cal K} & 0 \end{pmatrix},
\label{matrix1}
\\
 \boldsymbol{\eta} &= \begin{pmatrix} \eta_1 & \eta_2 \\ \eta_2 & \eta_3 \end{pmatrix},\quad&
 \boldsymbol{\zeta} &= \begin{pmatrix} \zeta_1 & \zeta_2 \\ \zeta_2 & \zeta_3 \end{pmatrix},
\label{matrix2}
\end{alignat}
and $\boldsymbol{\eta}$ and $\boldsymbol{\zeta}$ are positive semidefinite.

From Eqs.~(\ref{require1}) and (\ref{require2}), the energy-momentum tensor is determined as
\begin{align}
 \cT^{MN} &= 2\,(\mathsf{e}+\mathsf{p})\,\bigl(U^{\ubar{M}}\,U^{\bar{N}} + U^{\bar{M}}\,U^{\ubar{N}}\bigr) + \mathsf{p}\,\cH^{MN} \nn \\
 &\quad + \hat{\tau}^{MN} + \tau\,\Delta^{MN}\,,
\\
 \hat{\tau}^{MN} &= 2\,(\eta_2-\cG)\,\cE_{MN} + \eta_3 \, \gLie_U\cH_{MN} \,,
\\
 \tau &= (\zeta_2-{\cal K})\,\varepsilon + 2\,\zeta_3 \,\gLie_U d \,,
\end{align}
and the time evolution of the strain tensors is given by
\begin{align}
\label{eq:rheology1}
 [\gLie_U\cE]_{MN} & = -\frac{1}{\tau_1} \,\cE_{MN} - \frac{\cG+\eta_2}{2\,\lambda}\, \gLie_U\cH_{MN}\,, \\ 
 \gLie_U\varepsilon & = -\frac{1}{\tau_0} \,\varepsilon - \frac{2\,({\cal K}+\zeta_2)}{\gamma}\,\gLie_U d\,,
\label{eq:rheology2}
\end{align}
where
\begin{align}
 \tau_1 = \frac{\lambda}{\eta_1} \,, \quad \tau_0 = \frac{\gamma}{\zeta_1} \,.
\end{align}
Under these relations, the entropy never decreases. Eqs.~\eqref{eq:rheology1} and \eqref{eq:rheology2}, which determine the time evolution of the strain, are the equivalent to the so-called rheology equations and the parameters $\tau_1$ and $\tau_0$ correspond to relaxation times. At the time scale which is longer than $\tau_1$ and $\tau_0$\,, the temporal derivatives such as $\gLie_U \cE_{MN}$ and $\gLie_U \varepsilon$\,, are much smaller than $\frac{1}{\tau_1}\,\cE_{MN}$ and $\frac{1}{\tau_0}\,\varepsilon$ and we obtain
\begin{align}
 \cE_{MN} \propto \gLie_{U}\cH_{MN}\,,\qquad
 \varepsilon \propto \gLie_{U}d\,.
\end{align}
Then ${\cal T}_{M N}$ reduces to the energy-momentum tensor of viscous fluids,
\begin{align}
 \cT_{MN} & = 2\,(\mathsf{e}+\mathsf{p}) \,\bigl(U_{\bar{M}}\,U_{\ubar{N}} + U_{\ubar{M}}\,U_{\bar{N}}\bigr) + \mathsf{p}\,\cH_{MN} \nn \\
 &\quad + \eta\, \gLie_{U}\cH_{MN} + 2\,\zeta\,\gLie_{U}d\,\Delta_{MN} \,,
\label{newT}
\end{align}
where
\begin{align}
 \eta = \frac{\cG^2 + \det \boldsymbol{\eta}}{\eta_1} \,,\qquad 
 \zeta = \frac{{\cal K}^2 + \det \boldsymbol{\zeta}}{\zeta_1} \,.
\end{align}

\subsection{The general non-dissipative case}

Now we observe an interesting aspect of the model proposed in the previous Subsection. If we set $\boldsymbol{\eta}=0$ and $\boldsymbol{\zeta}=0$ in (\ref{matrix1})--(\ref{matrix2}), we obtain 
\begin{align}
 \cT^{MN} &= 2\,\mathsf{e}\,\bigl(U^{\ubar{M}}\,U^{\bar{N}} + U^{\bar{M}}\,U^{\ubar{N}}\bigr)  - 2\, \cG \,\cE^{MN}\nn \\
 &\quad + (\mathsf{p}-{\cal K}\,\varepsilon) \,\Delta^{MN} \,.
\end{align}
In this particular case, the Lie derivative for the generalized strains is given by
\begin{align}
 {} [\gLie_U\cE]_{MN} & = - \frac{\cG}{2\,\lambda}\, \gLie_U\cH_{MN}\,, 
\\ 
 \gLie_U\varepsilon & = - \frac{2\,{\cal K}}{\gamma}\,\gLie_U d\,.
\end{align}
Under $\boldsymbol{\eta}=0$ and $\boldsymbol{\zeta}=0$\,, the right-hand side of (\ref{EntropyProd}) vanishes and the entropy is conserved, $\nabla_M S^M=0$\,. This suggests that even the description of the perfect fluid in the double space could be enriched by the inclusion of new degrees of freedom to (\ref{proposaltmn}), as the generalized strain tensor. The energy-momentum tensor (\ref{proposaltmn}) obtained by the generalized scalar-perfect fluid identification is a particular case of the non-dissipative model described in this Section.
It is noted that, if we choose $\cG=\lambda$ and ${\cal K}=\gamma$, the material does not undergo a plastic deformation at the supergravity level and can be identified as the elastic material in the sense discussed in Section 3.3 of \cite{Y2}. 

So far we have constructed a proposal to incorporate imperfect fluid contributions in the double space considering a generalized strain tensor. We show in the next Section the parametrization of this model. Finally, in Section \ref{duality}, we discuss the relation between the non-dissipative model here constructed and the observations given in \cite{GasVene} about the non-invariance of the perfect fluid under T-duality rotations.    

\section{Parametrization}
For convenience, here we rewrite our fluid model by using the standard supergravity fields. 
We start by parameterizing the generalized velocity, $U^M$, and $V^M$ as
\begin{align}
 U_M = \begin{pmatrix}
 \delta_m^p & B_{mp} \\
 0 & \delta_p^m
 \end{pmatrix} \begin{pmatrix} v_p \\ u^p
\end{pmatrix}, \ 
 V_M = \begin{pmatrix}
 \delta_m^p & B_{mp} \\
 0 & \delta_p^m
 \end{pmatrix} \begin{pmatrix} u_p \\ v^p
\end{pmatrix}. 
\end{align}
Since the generalized velocity must satisfy the constraint $U_M\,U^M=2\,v_m\,u^m=0$\,, we may choose $v_m=0$\, and in this case $\cH_{MN}\,U^M\,U^N=-1$ shows that the velocity field $u^m$ satisfies the standard relation
\begin{align}
 g_{mn}\,u^m\,u^n = -1\,.
\end{align}

In what follows we keep both $u^m$ and $v_m$ without assuming a particular solution to the previous constraint. The fundamental equation of thermodynamics becomes
\begin{align}
 T\,\delta\tilde{s}_0 = \delta\tilde{\mathsf{e}} + \tilde{\mathsf{p}}\,\delta \ln v -2\,\tilde{\mathsf{p}}\,\delta \phi + \tilde{\mathsf{p}}\,u^{m} v^{n}\,\delta B_{mn}\,,
\label{TdS}
\end{align}
where we have defined $\tilde{\mathsf{e}}= \Exp{-2d}\mathsf{e}$ and $\tilde{\mathsf{p}}= \Exp{-2d}\mathsf{p}$, and $\delta \ln v = \frac{1}{2}\,\Delta^{mn}\, \delta g_{mn}$ with $\Delta^{mn} = u^m\,u^n - v^m\,v^n+g^{mn}$ corresponds to the variation of the spatial volume. 
The generalized Lie derivatives of the generalized metric and dilaton along the flow become
\begin{align}
 &\gLie_U\cH_{MN} 
\\
 &=\begin{pmatrix}
 \delta_m^p & B_{mp} \\
 0 & \delta_p^m
 \end{pmatrix}
 \begin{pmatrix}
 k_{(pq)} & g_{pr}\,k^{[rq]} \\
 -k^{[pr]}\,g_{rq} & -k^{(pq)}
 \end{pmatrix} 
 \begin{pmatrix}
 \delta_n^q & 0 \\
 -B_{qn} & \delta_q^n
 \end{pmatrix},
\nn\\
 &\gLie_U d = u^m\,\partial_m \phi - \tfrac{1}{2}\,\covD_m u^m \,,
\end{align}
where
\begin{align}
 k^{(mn)} & = 2\,\covD^{(m} u^{n)} \,, \\
 k^{[mn]} & = H^{mnr}\,u_r + 2 \covD^{[m} v^{n]}\,.
\end{align}
We also parameterize the strain tensor $\cE_{MN}$ in terms of a pair of symmetric and antisymmetric matrices $E_{m n}$ and $F^{m n}$, respectively, 
\begin{align}
 \cE_{MN} &=\begin{pmatrix}
 \delta_m^p & B_{mp} \\
 0 & \delta_p^m
 \end{pmatrix}
 \begin{pmatrix}
 - E_{pq} & - g_{pr}\,F^{rq} \\
 F^{pr}\,g_{rq} & E^{pq}
 \end{pmatrix} 
 \begin{pmatrix}
 \delta_n^q & 0 \\
 -B_{qn} & \delta_q^n
 \end{pmatrix},
\end{align}
and we parametrize the components of the generalized energy-momentum tensor (\ref{tmntotal}) and (\ref{tmnmixed}) as
\begin{align}
\!\!\!
 t^{(mn)} & = \bigl(\mathsf{e}-\tfrac{1}{2}\,\cT\bigr)\,(u^m\,u^n-v^m\,v^n)
\nn\\
 &\quad - 2\,(\cG-\eta_2)\,E^{mn} -2\,\eta_3\,\covD^{(m}u^{n)}\,, 
\\
\!\!\!
 t^{[mn]} & = - 2\,(\cG-\eta_2)\,F^{mn} - \eta_3\,H^{mnp}\,u_p
\nn\\
 &\quad +2\,\bigl(\mathsf{e}-\tfrac{1}{2}\,\cT\bigr)\,u^{[m}\,v^{n]} -2\,\eta_3\,\covD^{[m}v^{n]}\,,
\\
\!\!\!
\cT & = -2\,\mathsf{p} + 2\,({\mathcal K}-\zeta_2)\,\varepsilon + 2\,\zeta_3 \,\bigl(\covD_m u^m - 2\,\dot{\phi}\bigr)\,,
\end{align}
where $\dot{\phi}=u^m\,\partial_m \phi$. The previous expressions can be rewriten in terms of the ordinary sources as 
\begin{align}
\label{newTmn}
 T^{mn} &= \Exp{-2\phi}\,(t^{(mn)}-\tfrac{1}{2}\,\cT\,g^{mn}) \nn \\
 &= (e+p)\,(u^m\,u^n-v^m\,v^n) + p\,g^{mn} \nn \\
 &\quad - 2\,(\hat{\cG}-\hat{\eta}_2)\,E^{mn}- (\hat{\cal K}-\hat{\zeta}_2)\,\varepsilon\,\Delta^{mn}  \nn \\
 &\quad -2\,\hat{\eta}_3\,\covD^{(m}u^{n)}- \hat{\zeta}_3 \,\bigl(\covD_p u^p - 2\,\dot{\phi} \bigr)\,\Delta^{mn}\,,
\\
 J^{mn} &= - (\hat{\cG}-\hat{\eta}_2)\,F^{mn} - \tfrac{1}{2}\,\hat{\eta}_3\,H^{mnp}\,u_p 
\nn\\
 &\quad + \bigl(e+\tfrac{1}{2}\,\sigma\bigr)\,u^{[m}\,v^{n]} -\hat{\eta}_3\,\covD^{[m}v^{n]}\,,
\\
 \sigma &= 2\,p - 2\,(\hat{\mathcal K}-\hat{\zeta}_2)\,\varepsilon - 2\,\hat{\zeta}_3 \,\bigl(\covD_m u^m - 2\,\dot{\phi} \bigr)\,,
\end{align}
where the hatted quantities contains an additional dilaton factor, e.g., $\hat{\cG}=\Exp{-2\phi}\,\cG$\,. The rheology equations become
\begin{align}
 \Lie_u E^{mn} & = -\frac{1}{\tau_1} \,E^{mn} + \frac{\cG+\eta_2}{\lambda}\,\covD^{(m}u^{n)} \, ,
\\
 \Lie_u F^m{}_{n} & = -\frac{1}{\tau_1} \,F^m{}_{n} + \frac{\cG+\eta_2}{2\,\lambda}\,H^m{}_{np}\,u^p \, ,
\\
 \Lie_u\varepsilon & = -\frac{1}{\tau_0} \,\varepsilon + \frac{{\cal K}+\zeta_2}{\gamma}\, \bigl(\covD_m u^m - 2\, \dot{\phi} \bigr)\,, 
\end{align}
where we have ignored the quadratic dependence on the strain tensors and $\covD_{(m}u_{n)}$. Finally, in the fluid limit, the energy-momentum tensor becomes
\begin{align}
 T^{mn} & = (e+p)\,(u^m\,u^n-v^m\,v^n) + p\,g^{mn} -2\,\hat{\eta}\,\covD^{(m}u^{n)} \nn\\
 &\quad - \hat{\zeta} \,\bigl(\covD_p u^p - 2\, \dot{\phi} \bigr)\,\Delta^{mn}\,.
\end{align}

So far we have shown the form of the imperfect contributions in the double space after solving the strong constraint and imposing a parametrization for the new degrees of freedom. One important advantage of this new model is that now one can rewrite in a fully $\OO(D,D)$-covariant way the whole family of cosmological scenarios related to the perfect fluid dynamics in the supergravity approach. We use the next Section to discuss this point. 

\section{Comments on T-duality and perfect/imperfect fluids}
\label{duality}

The generalized energy-momentum tensor is a multiplet of the duality group and therefore its transformation under this symmetry is given by
\begin{align}
 \cT_{M N} \rightarrow h_{M}{}^{P}\, \cT_{P Q}\,h^{Q}{}_{N} \,,
\end{align}
where $h_{M}{}^{N}$ is an element of the $\OO(D,D)$ group. 

In \cite{GasVene}, a particular configuration of perfect fluids was considered in a cosmological background in $1+2$ dimensions without dilaton or $B$-field sources. Their original configuration can be summarized as
\begin{align}
 &g_{00}=-1,\quad g_{ij} = a^2(t)\,\delta_{ij},\quad B_{mn}=0\,,\nn
\\
 &T^m{}_n = {\rm diag}(-\rho,p_{\rm eff}\,\delta^i{}_j),\quad J^{mn} = \sigma=0 \,,\label{cosmobackground}
\end{align}
where $u^m=(1,0,0)$ and $v_m=0$. It was pointed out that if we perform an $\OO(D-1,D-1)$ transformation, the spatial energy-momentum tensor of the perfect fluid in (\ref{cosmobackground}) is mapped to the one with imperfect non-diagonal terms, thus the standard formulation of perfect fluids is not $\OO(D-1,D-1)$ covariant unless $p_{\rm eff}=0$.

Interestingly enough, it is not possible to reproduce the matter configuration (\ref{cosmobackground}) from the generalized energy-momentum tensor of the perfect fluid-scalar field correspondence (\ref{Gemt}) with vanishing dilaton charge $\sigma=0$ and vanishing $B$-field source $J^{mn}=0$ (see \cite{EN2}). In fact the parametrized energy-momentum tensor coming from (\ref{Gemt}) reads
\begin{align}
T^m{}_n = {\rm diag}(-e,p\,\delta^i{}_j)\,.
\end{align}
with $p=\sigma/2$. Therefore for vanishing dilaton source we obtain $p=0$ and in turn we can only recover (\ref{cosmobackground}) with $p_{\rm eff}=0$. This implies that in order to encode the general configuration (\ref{cosmobackground}) in a duality invariant way, we need to include new degrees of freedom in (\ref{Gemt}) in addition to the generalized metric, the generalized dilaton and the generalized velocity. One possibility to solve this problem is to consider the general non-dissipative scenario proposed here setting $\boldsymbol{\eta}=0$ and $\boldsymbol{\zeta}=0$ in equation (\ref{newTmn}). Thus the spatial mixed part of the energy-momentum tensor reads
\begin{align}
T^i{}_j = \left(p-{\cal K}\,\varepsilon\right)\delta^i{}_j-2\,\cG\,E^{i}{}_{j}\,,
\end{align}
with $p-{\cal K}\,\varepsilon=\sigma/2$. Indeed if we further apply $\sigma=0$ and $E^{i}{}_{j}=-(p_{\rm eff}/2\cG)\,\delta^i{}_j$, we obtain
\begin{align}
p - {\cal K}\,\varepsilon=0 \qquad T^i{}_j = p_{\rm eff}\,\delta^i{}_j\,.
\end{align}
with the constraint
\begin{align}
 \Lie_u E^{i}{}_{j} = \frac{\cG}{2\,\lambda}\,\bigl(\covD^{i}u_{j} + \covD_j u^i \bigr) \,,
\end{align}
coming from the rheology equation. Therefore from these expressions we can now read an effective non-vanishing pressure $p_{\rm eff}$ even for the case $\sigma=0$.
Since our model is duality covariant, we can find the T-dual configuration, which also contains the non-vanishing elastic strain, and there is no issue of the non-covariance. 

\section{Conclusions and outlook}\label{Conclusions}

In this work we have constructed the generalization of the Maxwell-Juttner distribution function (\ref{MBDFT}) as a solution of the generalized Boltzmann equation (\ref{BoltzmannDFT}). This function opens several possibilities in the double space. On one hand, all the conserved quantities (${\cal T}_{M N}$ and $S_{M}$) can be exactly constructed through integration in the generalized momentum coordinates in (\ref{EMDFT}) and (\ref{SDFT}). While this integration is a challenging task, it could serve as a crucial test to compare with the proposal formulated in \cite{EN2} through the generalized scalar-perfect fluid correspondence \cite{EN3}.

On the other hand, the main result of this work is the construction of a simple model to include non-equilibrium contributions (\ref{newT}). While the new terms in ${\cal T}_{MN}$ are fully covariant under generalized diffeomorphisms, it would be interesting to construct these contributions from the (double) statistical approach outlined in \cite{EN1}. In this sense the new terms should appear as a particular perturbation of the distribution function, taking into account collision contributions from the RHS of the generalized Boltzmann equation. 

Furthermore, we have proposed a way to measure the entropy production in a given system considering the relation (\ref{EntropyProd}). This expression can be used to define the second law of thermodynamics in the double space and the additional terms in the generalized energy-momentum tensor, after parametrization, introduce modifications to the Navier-Stokes fluids that changes the dynamics at a short-time scale as showed in \cite{Y1}--\cite{Y2}. In the proposed model, by choosing $\boldsymbol{\eta}=0$ and $\boldsymbol{\zeta}=0$, one can establish a model where the entropy current is conserved but the effect of the elasticity is included. Through this model we have explained how to recover the energy-momentum tensor of the perfect fluid in an $\OO(D,D)$-covariant way. 

\subsection*{Acknowledgements}
E.L is grateful to the organizers of the MITP scientific program “Higher Structures, Gravity and Fields” and to G. Itsios for discussions in the initial stage of this work. E.L and Y.S are very grateful to the organizers of 'Gravity beyond Riemannian Paradigm' (CQUesT-APCTP) where part of the work was developed.  E.L is supported by the SONATA BIS grant
2021/42/E/ST2/00304 from the National Science Centre (NCN), Poland. N.M.G is supported by CONICET PIP 11220170100817CO grant. Y.S is supported by JSPS KAKENHI Grant Number JP23K03391.

\end{document}